\documentstyle[12pt]{article}
\begin{document}

\begin{center}
{\Large \bf Black hole motion in Euclidean space as \linebreak a diffusion process II} \\
\vskip .5cm
K. Ropotenko\\
\centerline{\it State Service for Special Communication and}
\centerline{\it Information Protection of Ukraine,} \centerline{\it
13 Solomianska str., Kyiv, 03680, Ukraine}
\bigskip
\verb"ropotenko@ukr.net"

\end{center}
\bigskip\bigskip
\begin{abstract}
A diffusion equation approach to black hole thermodynamics in
Euclidean sector is proposed. A diffusion equation for a generic
Kerr-Newman black hole in Euclidean sector is derived from the Bloch
equation. Black hole thermodynamics is also derived and it is found,
in particular, that the entropy of a generic Kerr-Newman black hole
is the same, apart from the logarithmic corrections, as the
Bekenstein-Hawking entropy of the black hole.
\end{abstract}
\bigskip\bigskip

In \cite{ro}, I derived a diffusion equation for a Schwarzschild
black hole from the Bunster-Carlip equations and showed that the
black hole evolution in Euclidean sector exhibits a diffusion
process. Namely I showed that the Bunster-Carlip equations
\begin{equation}
\label{eq1} \frac{\hbar}{i}\frac{\partial \psi}{\partial t}+M\psi=0,
\end{equation}
\begin{equation}
\label{eq2} \frac{\hbar}{i}\frac{\partial \psi}{\partial
\Theta}-\frac{A}{8\pi G}\psi=0,
\end{equation}
where $t$ is the lapse of asymptotic proper time at spatial infinity
and $\Theta$ is the lapse of the hyperbolic angle at the horizon,
transform to the equation
\begin{equation}
\label{eq3} \frac{\partial \rho}{\partial \Theta_{\rm
E}}=D\frac{\partial^{2}\rho}{\partial t_{\rm E}^{2}}
\end{equation}
for the probability density $\rho=|\psi(t_{\rm E}$, $\Theta_{\rm
E})|^{2}$ under Euclidean continuation $\Theta_{\rm E}=i\Theta$ and
$t_{\rm E}=it$. In general, the Euclidean time coordinate $t_{\rm
E}$ behaves like a spatial coordinate. But (\ref{eq3}) is an
one-dimensional diffusion equation in the temporal $\Theta_{\rm E}$
and spatial $t_{\rm E}$ coordinates with the diffusion coefficient
$D=2G\hbar$. So I reinterpreted the Euclidean time coordinate
$t_{\rm E}$ as a spatial coordinate and $\Theta_{\rm E}$ as a
temporal coordinate. In place of $\Theta_{\rm E}$ and $t_{\rm E}$ I
will hereafter write $\Theta$ and $\tau$. After analytical
continuation to the cyclic imaginary time, we deal with a quantum
system at a finite temperature, so the black hole should be
described not by the probability density $\rho (x)$ but by the
canonical density matrix $\rho (x,x^{\prime};\beta)$. Determining
$\rho (x,x^{\prime};\beta)$ I found that the entropy of a
Schwarzschild black hole is the same, apart from the logarithmic
corrections, as the Bekenstein-Hawking entropy.

In the literature, the path integral approach is the only approach
to black hole thermodynamics in Euclidean quantum gravity. In this
note I propose an alternative diffusion equation approach. I derive
a diffusion equation for the density matrix of a generic Kerr-Newman
black hole in Euclidean sector immediately from the Bloch equation.
I also derive black hole thermodynamics and find, in particular,
that the entropy of a Kerr-Newman black hole is the same, apart from
the logarithmic corrections, as the Bekenstein-Hawking entropy of
the black hole.

I begin with the Bloch equation \cite{fey}
\begin{equation}
\label{eq4} \frac{\partial \rho}{\partial \beta}=-H \rho
\end{equation}
for a Schwarzschild black hole at a temperature $T_{\rm
H}=\beta^{-1}$.  The (unnormalized) canonical density matrix has the
form $\rho(\beta)= e^{-\beta H}$. Note that here, as in ordinary
statistics, the form of the density matrix must be regarded only as
a postulate, to be justified solely on the basis of agreement of its
predictions with the thermodynamical properties of black holes. As
before, I adopt the Euclidean approach and consider the black hole
motion in Euclidean sector in the temporal $\Theta$ and spatial
$\tau$ coordinates, so that $\rho(\beta)= \rho
(\tau,\tau^{\prime};\beta)$ in coordinate representation. For
physical applications $\Theta$ should be set equal to $2\pi$, so
that $\beta =2\pi/k$, where $k$ is the surface gravity. But for the
purposes of thermodynamic analysis I shall keep $\Theta$ arbitrary
and put $\Theta=2\pi$ only in some final results. Therefore,
(\ref{eq4}) reads
\begin{equation}
\label{eq5} \frac{\partial \rho}{\partial \Theta}=-\frac{1}{k}H
\rho.
\end{equation}
Next, I define the Hamiltonian of a black hole with the mass $M$ as
that of a free particle moving along the coordinate $\tau$
\begin{equation}
\label{eq6} H=\frac{p^{2}}{2M},
\end{equation}
and obtain
\begin{equation}
\label{eq7} \frac{\partial \rho}{\partial
\Theta}=D\frac{\partial^{2}\rho}{\partial \tau^{2}},
\end{equation}
where $D=\hbar/2kM$. This is an one-dimensional diffusion equation,
and we can write down its solution readily:
\begin{equation}
\label{eq8} \rho (\tau,\tau^{\prime};\Theta) =\frac{1}{\sqrt{4\pi D
\Theta}}\exp{\left[{-\frac{(\tau-\tau^{\prime})^{2}}{4D\Theta}}\right
]},
\end{equation}
where the proportionality factor is chosen such that
\begin{equation}
\label{eq9} \rho (\tau,\tau^{\prime};0)=\delta(\tau-\tau^{\prime}).
\end{equation}
The crucial property of the black hole solutions in Euclidean sector
is their periodicity in the imaginary time, $\tau \sim \tau+\beta$,
where $\beta=2\pi/k$. Therefore
\begin{equation}
\label{eq10} \rho (\tau,\tau;\Theta)=\frac{1}{\sqrt{4\pi D
\Theta_{\rm E}}}\exp{\left({-\frac{\beta^{2}}{4D\Theta}}\right )}.
\end{equation}
For a linear system of length $\beta=2\pi/k$, the integration over
$\tau$ gives
\begin{equation}
\label{eq11} Z(\beta)=\int^{\beta}_0 \rho (\tau,\tau) d\tau =
\frac{\beta}{\sqrt{4\pi D
\Theta}}\exp{\left({-\frac{\beta^{2}}{4D\Theta}}\right )}.
\end{equation}
This is the partition function for a Schwarzschild black hole.
Differentiating with respect to the $\beta$ and then putting
$\Theta=2\pi$, we obtain the internal energy
\begin{equation}
\label{eq12} E=-\frac{\partial \ln Z(\beta)}{\partial
\beta}=M-\beta^{-1}.
\end{equation}
It is the same, apart from the Hawking temperature, as the mass of
the black hole. The entropy of the black hole is given by
\begin{equation}
\label{eq13} S=\ln Z +\beta E =\frac{A}{4l_{\rm
P}^{2}}+\frac{1}{2}\ln \left(\frac{A}{4l_{\rm
P}^{2}}\right)+\ln\left(\frac{1}{e\sqrt{\pi}}\right).
\end{equation}
It is the same, apart from the logarithmic corrections, as the
Bekenstein-Hawking entropy $S_{\rm BH}=A/4l_{\rm P}^{2}$.

Let us now consider a Kerr-Newman black hole at a temperature
$T_{\rm H}=\beta^{-1}$. As is well known, in the near-horizon
approximation the Euclidean sector of a Kerr-Newman black hole is
similar to that of a Schwarzschild black hole. So, as in the case of
a Schwarzschild black hole, one can introduce the corresponding
temporal $\Theta$ and spatial $\tau$ coordinates for a Kerr-Newman
black hole. I define the Hamiltonian of a Kerr-Newman black hole as
that of a particle moving in the potential $U=\Omega J +
\frac{1}{2}\Phi Q$,
\begin{equation}
\label{eq14} H=\frac{p^{2}}{2M}+\Omega J + \frac{1}{2}\Phi Q,
\end{equation}
where all quantities have the standard meaning. Proceeding exactly
as in the derivation of (\ref{eq7}), we get the equation
\begin{equation}
\label{eq15} \frac{\partial \rho}{\partial
\Theta}=D\frac{\partial^{2}\rho}{\partial \tau^{2}}-b\rho,
\end{equation}
where $D=\hbar/2kM$ and
\begin{equation}
\label{eq16}b=\frac{1}{k}\left(\Omega J +\frac{1}{2}\Phi Q\right ).
\end{equation}
This is also a diffusion equation but with the convection term
$-b\rho$. The equation describes the probability flow in Euclidean
sector of a Kerr-Newman black hole with both diffusion
$D\partial^{2}\rho/\partial \tau^{2}$ along the axis $\tau$ and the
outflow $-b\rho$ in a direction perpendicular to the axis. How much
diffusion and convection takes place depends on the relative size of
the two coefficients $D$ and $b$. The solution to the equation (at
the initial condition (\ref{eq9})) is
\begin{equation}
\label{eq17} \rho=\frac{1}{\sqrt{4\pi D
\Theta}}\exp{\left[{-\frac{(\tau-\tau^{\prime})^{2}}{4D\Theta}}\right
]}\exp (-b\Theta).
\end{equation}
There is also an integral formula for solutions to (\ref{eq15}),
known as the Feynman-Kac formula; it is an integral over path space
with respect to Wiener measure \cite{fey}. Since, as in the
Schwarzschild case, $\tau \sim \tau+\beta$, where $\beta=2\pi/k$,
the trace of (\ref{eq17}) leads to
\begin{equation}
\label{eq18} Z(\beta)=\frac{\beta}{\sqrt{4\pi D
\Theta}}\exp{\left({-\frac{\beta^{2}}{4D\Theta}}\right )}\exp
(-b\Theta).
\end{equation}
This this the partition function for a Kerr-Newman black hole.
Differentiating with respect to the $\beta$ and then putting
$\Theta=2\pi$, we obtain the internal energy
\begin{equation}
\label{eq19} E=-\frac{\partial \ln Z_g(\beta)}{\partial
\beta}=M-\beta^{-1}.
\end{equation}
It is the same, apart from the Hawking temperature, as the mass of
the black hole. Finally, the entropy of a Kerr-Newman black hole is
given by
\begin{eqnarray}
\lefteqn{S=\ln Z +\beta E =\frac{A}{4l_{\rm P}^{2}}+\frac{1}{2}\ln
\left(\frac{A}{4l_{\rm P}^{2}}\right)+{}} \nonumber \\
&& {}+\frac{1}{2}\ln\left(\frac{M}{\sqrt{M^{2}-a^{2}-Q^{2}}}\right)
+\ln\left(\frac{1}{e\sqrt{\pi}}\right).
\end{eqnarray}
It is the same, apart from the logarithmic corrections, as the
Bekenstein-Hawking entropy of the black hole.

The statistical interpretation of the Bekenstein-Hawking entropy
remains a central problem in black hole physics. The path integral
approach in Euclidean quantum gravity cannot answer what the degrees
of freedom are responsible for the entropy. In this note, the black
hole motion in Euclidean sector is modeled as an one-dimensional
diffusion in the coordinates $\tau$ and $\Theta$. A random walk
process is the basis of the diffusion. Thus, if the model is
correct, then there should exist the black hole constituents with
the Boolean degrees of freedom $\pm\Theta$ or $\pm \tau$.

\end{document}